%% file: main.tex
\documentclass[10pt,conference]{IEEEtran}
\IEEEoverridecommandlockouts

\usepackage{multirow}
\usepackage{array, tabularx}
\usepackage{booktabs}
\usepackage{subcaption}
\usepackage{float}
\usepackage{enumitem}
\usepackage{balance}

\usepackage{adjustbox}
\usepackage{caption}
\usepackage{url}
\usepackage{amsmath,amssymb,amsfonts}
\usepackage{algorithmic}
\usepackage{graphicx}
\usepackage{textcomp}
\usepackage{xcolor}
\usepackage{cite}

\usepackage[colorlinks=true, urlcolor=blue, citecolor=black, linkcolor=black]{hyperref}

\usepackage[most]{tcolorbox}
\usepackage{csquotes}
\newtcolorbox{myquote}[1][]{%
    colback=black!5,
    colframe=black!5,
    notitle,
    sharp corners,
    borderline west={2pt}{0pt}{black!80!black},
    enhanced,
    breakable,
    top=0.5pt,
    bottom=0.5pt
}

\usepackage{acronym}
\usepackage{glossaries}
\input{acronyms} %% Acronyms Management

\def\BibTeX{{\rm B\kern-.05em{\sc i\kern-.025em b}\kern-.08em
    T\kern-.1667em\lower.7ex\hbox{E}\kern-.125emX}}
\begin{document}

\title{A Large-Scale Study on the Development and Issues of Multi-Agent AI Systems}

\newcommand{\stitle}[1]
{\noindent\textup{\textbf{#1}}}
\newcommand\myNum[1]{\emph{{(#1)}}}

\newcommand\todou[1]{\textcolor{red}{\textit{Umar: #1}}}
\newcommand\todok[1]{\textcolor{blue}{\textit{Krishna: #1}}}
\newcommand\todov[1]{\textcolor{blue}{\textit{Vinaik: #1}}}

\author{\IEEEauthorblockN{Daniel Liu\IEEEauthorrefmark{1}, Krishna Upadhyay\IEEEauthorrefmark{1}, Vinaik Chhetri\IEEEauthorrefmark{1}, A.B. Siddique\IEEEauthorrefmark{2}, Umar Farooq\IEEEauthorrefmark{1}}
\IEEEauthorblockA{\IEEEauthorrefmark{1}
    Louisiana State University, 
    \IEEEauthorrefmark{2} 
    University of Kentucky \\
 Email: 
dliu26@lsu.edu, kupadh4@lsu.edu, vchhet2@lsu.edu, ab.siddique@uky.edu, ufarooq@lsu.edu
}}

\maketitle

\thispagestyle{plain}
\pagestyle{plain}
\begin{abstract}
The rapid emergence of multi-agent AI systems (MAS), including LangChain, CrewAI, and AutoGen, has shaped how large language model (LLM) applications are developed and orchestrated. 
However, little is known about how these systems evolve and are maintained in practice. This paper presents the first large-scale empirical study of open-source MAS, analyzing over $42K$ unique commits and over $4.7K$ resolved issues across eight leading systems. 
Our analysis identifies three distinct development profiles: sustained, steady, and burst-driven. These profiles reflect substantial variation in ecosystem maturity. Perfective commits constitute 40.8\% of all changes, suggesting that feature enhancement is prioritized over corrective maintenance~(27.4\%) and adaptive updates~(24.3\%). 
Data about issues shows that the most frequent concerns involve bugs~(22\%), infrastructure~(14\%), and agent coordination challenges~(10\%). 
Issue reporting also increased sharply across all frameworks starting in 2023. 
Median resolution times range from under one day to about two weeks, with distributions skewed toward fast responses but a minority of issues requiring extended attention.
These results highlight both the momentum and the fragility of the current ecosystem, emphasizing the need for improved testing infrastructure, documentation quality, and maintenance practices to ensure long-term reliability and sustainability.
\end{abstract}


\input{intro}

\input{background}

\input{study-settings}
\input{results}
\input{threats}

\section{Conclusions}

This study presents the first large-scale empirical analysis of development and maintenance practices in multi-agent AI systems. We examine eight of the most prominent open-source projects, covering over $42K$ unique commits and $4.7K$ resolved issues. By analyzing commit activity, code evolution, and issue handling, we offer a quantitative perspective on how these frameworks are built and sustained within open-source communities.
Our findings reveal an ecosystem undergoing rapid growth but still in the process of stabilizing. Development is largely driven by feature enhancement, with perfective commits comprising the majority of changes. Issue data indicate that bugs, infrastructure concerns, and agent coordination challenges are the most common problems reported. Resolution times vary widely across projects, reflecting differences in maturity and maintainer capacity. A sharp rise in commit and issue activity after 2023 signals increased adoption and community engagement, fueled by the rising use of \glspl*{llm} in \glspl*{mas}.

\bibliographystyle{plain}

\bibliography{bibliography.bib}

\end{document}

%% file: acronyms.tex
\newacronym{ai}{AI}{Artificial Intelligence}
\newacronym{llm}{LLM}{large language model}
\newacronym[shortplural=MAS,longplural=multi-agent systems]{mas}{MAS}{multi-agent system}
\newacronym{gqm}{GQM}{Goal Question Metric}
\newacronym{rq}{RQ}{research question}
\newacronym{lcel}{LCEL}{LangChain Expression Language}
\newacronym{rag}{RAG}{Retrieval-Augmented Generation}
\newacronym{pr}{PR}{pull request}
\newacronym[shortplural=CV,longplural=coefficients of variation]{cv}{CV}{coefficient of variation}
\newacronym{cli}{CLI}{command line interface}
\newacronym{ux}{UX}{user experience}
\newacronym[shortplural=IQR,longplural=interquartile ranges]{iqr}{IQR}{interquartile range}
\newacronym{api}{API}{application programming interface}

%% file: intro.tex
\section{Introduction}
\label{sec:intro}
The emergence of \glspl*{llm} has significantly influenced the design of intelligent systems, giving rise to \glspl*{mas} that support collaboration among autonomous agents on complex tasks. Frameworks such as AutoGen~\cite{autogen}, CrewAI~\cite{CrewAI}, and LangChain~\cite{LangChain} provide abstractions for agent creation, communication, and coordination. These systems support the construction of workflows that integrate reasoning, planning, and tool use, making them a central component of modern \gls*{llm}-based software development.

Despite their growing adoption, little is known about how these systems evolve and are maintained. Prior work on \glspl*{mas} has focused primarily on algorithmic and architectural advances, while the empirical understanding of their software development practices remains limited. This gap is critical because these systems must evolve rapidly to support new models, APIs, and orchestration mechanisms, while maintaining stability and long-term sustainability.

\begin{figure}[t]
    \centering
    \includegraphics{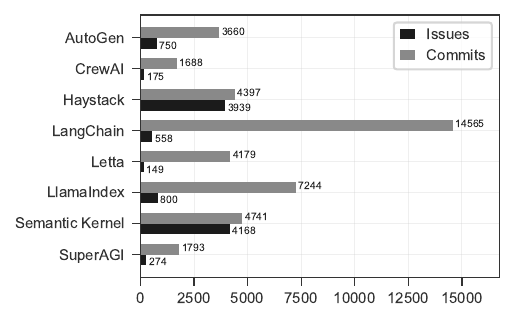}
    \vspace{-1em}
    \caption{Development and maintenance activity across eight major multi-agent AI systems, illustrating the commits and issues used in our large-scale study.}
    \label{fig:issues-commits-repository}
    \vspace{-2em}
\end{figure}

In software engineering, mining software repositories has proven effective for identifying development and maintenance trends at scale. However, such analyses have not yet been applied to multi-agent AI systems, which involve a mix of neural and symbolic reasoning, distributed execution, and frequent integration with external services. Understanding how these systems evolve and how issues are handled in practice can provide valuable insights into the engineering of complex AI systems. 

\begin{table*}
    \centering
    \caption{Overview of the eight multi-agent AI systems analyzed in this study, including their architecture types, primary purposes, and GitHub popularity.}
    \renewcommand{\arraystretch}{1.2}
    \footnotesize
    \begin{tabularx}{\textwidth}{ccXc}
        \toprule
        \textbf{MAS} & \textbf{Architecture} & \textbf{Purpose} & \textbf{GitHub Stars} \\
        \midrule
         Autogen~\cite{autogen} & Conversational Workflow & Multi-agent collaboration through automated conversational workflows with support for sequential, group, and nested chat patterns & 51.3K \\
         CrewAI~\cite{CrewAI} & Role-based Hierarchy & Orchestrating autonomous agents with defined roles working collaboratively on tasks through sequential or manager-led execution & 40K \\
         Haystack~\cite{haystack} & Graph-based Pipeline & Building Retrieval-Augmented Generation (RAG) applications, document search, question-answering, and agentic systems with modular, composable components & 23.2K \\
         LangChain~\cite{LangChain} & Modular Chain & General-purpose LLM orchestration enabling chains, agents, and tool integrations through composable building blocks & 119K \\
         Letta~\cite{Letta} & Memory-centric Agent & Building stateful AI agents with persistent memory, enabling long-term context retention and personalized interactions across conversations & 19K \\
         Llama Index~\cite{llamaindex} & Data-centric Pipeline & Context-augmented LLM applications focusing on connecting LLMs to external data sources through RAG patterns & 45K \\
         Semantic Kernel~\cite{semantic-kernel} & Plugin-based Orchestrator & Enterprise AI integration bridging LLMs with existing code through plugins, enabling AI orchestration in business applications & 26.6K \\
         SuperAGI~\cite{SuperAGI} & Extensible Agent Toolkit & Building, deploying, and managing production-ready autonomous AI agents with concurrent execution and tool integration capabilities & 16.8K \\
         \bottomrule
    \end{tabularx}
    \label{tab:overview-mas}
    \vspace{-2em}
\end{table*}

To address this gap, we conduct a large-scale empirical study of the most active open-source multi-agent AI systems on GitHub (ranging from $16K$ to $119K$ stars). Our dataset includes eight representative projects: AutoGen~\cite{autogen}, CrewAI~\cite{CrewAI}, Haystack~\cite{haystack}, LangChain~\cite{LangChain}, Letta~\cite{Letta}, LlamaIndex~\cite{llamaindex}, Semantic Kernel~\cite{semantic-kernel}, and SuperAGI~\cite{SuperAGI}. Table~\ref{tab:overview-mas} provides their architecture and usage details. From these repositories, we collect 42,267 unique commits and 4,731 resolved issues, enabling a detailed examination of development patterns and maintenance activities.

This study is structured around two core \glspl*{rq}. The first asks how development patterns differ across \glspl*{mas}, aiming to characterize variations in activity levels, growth trajectories, and commit types. The second investigates issue reporting and resolution, focusing on the types of issues encountered, their frequency, and how efficiently they are addressed across projects. Together, these \glspl*{rq} provide a comprehensive view of the development and maintenance practices within the multi-agent AI ecosystem.

As shown in Figure~\ref{fig:issues-commits-repository}, the selected repositories vary significantly in both development activity and issue volume. LangChain exhibits the highest development intensity with over $14K$ commits, followed by LlamaIndex and Semantic Kernel with approximately $7K$ and $4.7K$ commits, respectively. Haystack and AutoGen demonstrate steady, long-term activity, while CrewAI, Letta, and SuperAGI contribute at smaller but consistent scales.
On the issue side, Semantic Kernel and Haystack each report more than $4K$ issues, indicating large and active developer communities.
This variation highlights a maturing ecosystem in which a few MAS account for most of the development and maintenance work.

Our findings identify three development profiles. \emph{Sustained} systems like Haystack show consistent growth and periodic refactoring. \emph{Steady} ones such as Semantic Kernel and LlamaIndex maintain balanced activity with moderate fluctuation. \emph{Burst-driven}, including SuperAGI, undergo intense but short-lived development. LangChain stands out with over $14K$ commits, reflecting its central influence. These profiles reflect differing maturity, contributor engagement, and stability.

Commit-level analysis shows that perfective commits dominate~(40.8\%), indicating a strong focus on feature enhancement over corrective~(27.4\%) and adaptive~(24.3\%) changes. Code churn patterns since 2023 show increasing deletions matching insertions, signaling a shift from expansion to architectural refinement.
Issue analysis reveals that bugs~(22\%), infrastructure~(14\%), and agent coordination~(10\%) are the most frequent concerns, highlighting the complexity of orchestrating modular agents. Median resolution times range from under a day to two weeks, but high variance points to uneven responsiveness and maintenance capacity.

This work provides the first quantitative analysis of the development and maintenance of multi-agent AI systems. It is characterized by fast iteration, feature-oriented development, and uneven maturity across systems. Together, these findings highlight both the momentum and fragility of the ecosystem.
To support future research in this domain, we release our curated dataset.

%% file: background.tex
\section{Background and Related Work}
\label{sec:background}
\subsection{Multi-Agent AI System}
The rise of \glspl*{llm} has fundamentally changed how AI systems are designed and built. Instead of relying on a single, all-purpose model, developers are increasingly turning to multi-agent architectures, where multiple specialized agents work together to solve complex problems. Each agent has its own role, capabilities, and objectives, and they collaborate through structured communication and coordination. This shift allows systems to reason more effectively and distribute work across agents, mimicking teamwork in human organizations.

Several open-source frameworks have emerged to make the construction of such \glspl*{mas} more accessible. AutoGen~\cite{autogen} provides an environment for building conversational agents that engage in multi-turn dialogues, enabling them to negotiate, plan, and cooperate to achieve shared objectives. CrewAI~\cite{CrewAI} takes inspiration from human organizations by modeling agents as members of a hierarchical ``crew'', each with specific roles and responsibilities, facilitating structured collaboration and delegation. LangChain~\cite{LangChain} offers composable building blocks or ``chains'', that allow developers to design complex reasoning workflows, while LlamaIndex~\cite{llamaindex} focuses on augmenting agents with data retrieval capabilities, enabling them to access and reason over external knowledge sources efficiently. Microsoft's Semantic Kernel~\cite{semantic-kernel} introduces an enterprise-grade orchestration layer that blends symbolic and neural reasoning. It enables developers to define reusable ``skills'' and connect LLM-based reasoning with conventional code execution. Together, these frameworks illustrate the growing sophistication of modern AI development, where systems are not just powered by a single intelligent model but by networks of coordinated agents that interact dynamically to achieve goals.

Despite differences in design philosophy, these frameworks share common architectural foundations. Most provide abstractions for agent definition (roles, personas, capabilities), inter-agent communication (message passing, shared memory, context sharing), task decomposition (planning and delegation), and execution control (sequential, parallel, or conditional orchestration). The choice of framework typically reflects trade-offs between flexibility, simplicity, and coordination complexity, balancing how much autonomy individual agents have against how tightly their interactions are managed.

\subsection{Related Work}

In software engineering, software repository mining has long been a crucial method for understanding how software is built and maintained at scale.
A study on language adoption trends~\cite{ ray2014large} have shown how ecosystem choices influence code quality.
Bug predication research has leveraged historical data to identify defect-prone components~\cite{kim2007predicting, rahman2013sample}.
Eyolfson et al.~\cite{eyolfson2011time} revealed that even the timing of code commits, such as late-night activity, can impact software reliability.
Domain-specific analyses have provided further insights, identifying recurring problems such as API misuse in mobile apps~\cite{li2018cid}, vulnerability patterns in blockchain-based smart contracts~\cite{qian2022smart}, and documentation drift detected through commit message analysis~\cite{buse2010automatically}.

Building on this tradition, recent work on multi-agent AI systems has focused largely on their architectural and algorithmic innovations. Surveys such as those by Guo et al.~\cite{guo2024large} and Xi et al.~\cite{xi2025rise} categorize advances in agent architectures, communication protocols, coordination mechanisms, and memory management.
Benchmarking efforts like AgentBench~\cite{liu2023agentbench} provide standardized tasks for evaluating agent collaboration, reasoning, and adaptability.
Other studies have begun to explore the social dynamics that emerge when multiple agents interact, including cooperation and coordination behaviors in simulated environments~\cite{Park2023GenerativeAgents}, as well as challenges related to maintaining consistent communication and shared goals in dialogue-based collaboration.

This work addresses this gap through a large-scale empirical analysis of GitHub repositories that focus on \glspl*{mas}. By examining their structures, evolution patterns, and development activities, we aim to capture the current landscape of multi-agent software engineering in practice. The study highlights emerging trends in framework design and maintenance challenges, offering new insights into how these complex systems are engineered and evolved. Our findings contribute to a clearer understanding of \gls*{mas} development and lay the groundwork for future research on improving tools and practices in this rapidly growing field.

%% file: study-settings.tex
\section{Methodology}
\label{sec:method}

\subsection{Study Objective and Research Questions}
This study investigates how multi-agent AI systems are developed and maintained within open-source communities. We analyze repositories that implement agent-based architectures to gain insight into their development patterns and maintenance practices.
Two main \glspl*{rq} arise from this goal.

\begin{myquote}
\textbf{RQ1}. \textit{How do development patterns vary among the \gls*{mas} repositories?}
\end{myquote}

\begin{itemize}
    \item RQ1.1: What are the distinct commit activity patterns across repositories?
    \item RQ1.2: What is the distribution of commit types in \gls*{mas} repositories?
\end{itemize}

\begin{myquote}
\textbf{RQ2}. \textit{What are the issue reporting patterns and resolution characteristics in \gls*{mas} frameworks?}
\end{myquote}

\begin{itemize}
    \item RQ2.1: How do issue patterns evolve over time across frameworks?
    \item RQ2.2: What types of issues are most prevalent across frameworks?
\end{itemize}

\subsection{Dataset Construction}

\subsubsection{Data Collection}

We selected eight popular open-source GitHub repositories with diverse architectural patterns, as detailed in Table~\ref{tab:overview-mas}. Using the GitHub GraphQL API~\cite{github-graphql-api}, we extracted a total of 10,813 closed issues across these repositories. In addition to the issues dataset, we compiled the commit history by cloning each repository and extracting all commit records. This process yielded an initial dataset of 44,041 commits spanning the development history of the selected repositories.

\begin{table}[t]
    \centering
    \caption{Overview of dataset composition.}
    \renewcommand{\arraystretch}{1.2}
    \setlength{\tabcolsep}{3pt}
    \begin{tabularx}{\columnwidth}{
        l
        *{4}{>{\raggedleft\arraybackslash}X}
        >{\raggedleft\arraybackslash}X
    }
        \toprule
        \multirow{2}{*}[-.6em]{\textbf{MAS}} & 
        \multicolumn{4}{c}{\textbf{Issues}} & 
        \multirow{2}{*}[-.6em]{\textbf{Commits}} \\
        \cmidrule(lr){2-5}
        & \textbf{Total} & \textbf{w/ PRs} & \textbf{ Labeled} & \textbf{w/ Both} & \\
        \midrule
        AutoGen & 750 & 709 & 560 & 530 & 3,660 \\
        CrewAI & 175 & 89 & 167 & 82 & 1,688 \\
        Haystack & 3,939 & 1,405 & 2,720 & 1,062 & 4,397 \\
        LangChain & 558 & 511 & 238 & 220 & 14,565 \\
        Letta & 149 & 135 & 75 & 67 & 4,179 \\
        LlamaIndex & 800 & 770 & 780 & 762 & 7,244 \\
        Semantic Kernel & 4,168 & 1,108 & 3,897 & 1,069 & 4,741 \\
        SuperAGI & 274 & 4 & 82 & 1 & 1,739 \\
        \bottomrule
    \end{tabularx}
    \label{tab:dataset-stat}
    \vspace{-2em}
\end{table}
\subsubsection{Preprocessing}

Our analysis focused specifically on issues whose resolution involved code modifications. We therefore filtered out issues that lacked associated \glspl*{pr}, reducing the dataset from 10,813 to 4,731 issues.
We also examined labeled issues to understand issue categorization patterns in these repositories. Of the closed issues, 8,519 contained labels, and when combined with the \gls*{pr} requirement, 3,793 labeled issues with associated \glspl*{pr} remained in our dataset.

For the commit dataset, we addressed data quality issues stemming from Git operations such as cherry-picking and rebasing, which can introduce duplicate commits with identical content but different commit hashes. Through systematic duplicate detection and removal, we refined the dataset from 44,041 to 42,267 unique commits, ensuring data integrity for subsequent code change analyses.

Different subsets of these preprocessed datasets were used for different analyses throughout this study, with the specific dataset selection determined by the \glspl*{rq} being addressed. The breakdown for each repository is presented in Table~\ref{tab:dataset-stat}.

%% file: results.tex
\section{Results}
\label{sec:results}

\subsection{Development and Contribution Patterns (RQ1)}

\begin{figure}[t]
    \centering
    \begin{subfigure}[t]{0.49\linewidth}
        \centering
        \includegraphics[width=\linewidth]{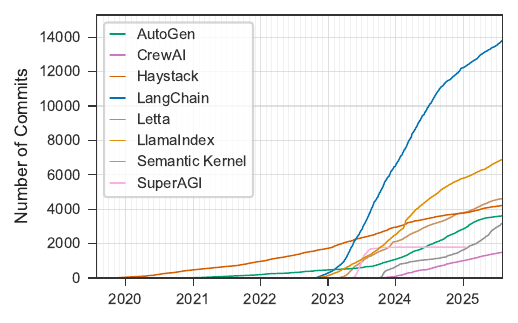}
        \label{fig:commits-cum-trend}
    \end{subfigure}
    \hfill
    \begin{subfigure}[t]{0.49\linewidth}
        \centering
        \includegraphics[width=\linewidth]{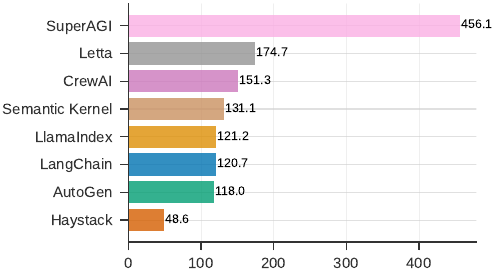}
        \label{fig:commits-cov-monthly}
    \end{subfigure}
    \vspace{-1.5em}
    \caption{Commit activity patterns across multi-agent AI frameworks. (a) Cumulative development growth. (b) Variation in monthly commit regularity, higher means irregular patterns.}
    \vspace{-2em}
    \label{fig:rq1-commit-trends}
\end{figure}

% RQ 1.1 >>>>>>>>>>
\subsubsection{What are the distinct commit activity patterns
across repositories?}
Commit activity across the analyzed \gls*{mas} repositories reveals three distinct development profiles. As shown in Figure~\ref{fig:rq1-commit-trends}~(a), LangChain leads with around 14,000 commits, reflecting rapid growth starting in mid-2023 and stabilization by 2025.

Haystack shows the most consistent trajectory, with steady contributions since 2020 and the lowest coefficient of variation (\gls*{cv}) at 48.6\%, as shown in Figure~\ref{fig:rq1-commit-trends}~(b). In contrast, SuperAGI follows a burst-driven pattern, marked by a sharp spike in mid-2023 and minimal activity afterward, resulting in a \gls*{cv}  of 456.1\%.

\begin{figure}[ht]
    \centering
    \vspace{-1.2em}\includegraphics{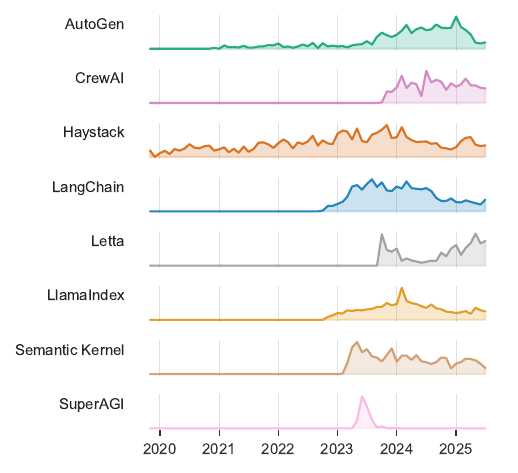}
    \vspace{-2em}
    \caption{Monthly commit activity sparklines for each repository showing temporal patterns.}
    \label{fig:commits-sparklines-monthly}
    \vspace{-1em}
\end{figure}

Other systems fall between these extremes, with patterns ranging from steady growth to sporadic surges. As Figure~\ref{fig:commits-sparklines-monthly} shows, most projects accelerated in 2023, marking a broader inflection point aligned with rising interest in AI agents.

\begin{figure}[t]
    \centering
    \includegraphics{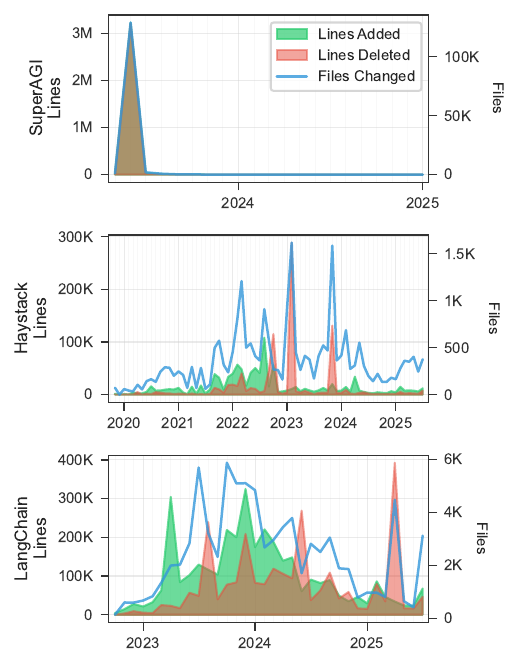}
    \vspace{-2em}
    \caption{Code churn patterns showing lines added, lines deleted, and files changed across different development profiles}
    \vspace{-1.5em}
    \label{fig:different-code-churn-patterns}
\end{figure}

To understand how the three development profiles translate into code evolution, we analyzed code churn patterns, which closely reflect the observed commit trends. SuperAGI added nearly 3 million lines in early 2023, followed by minimal maintenance, suggesting rapid prototyping with limited follow-up (Figure~\ref{fig:different-code-churn-patterns}). Haystack exhibits deliberate refactoring, marked by two large deletion events ($\sim140K$ and $\sim290K$ lines) in 2023 and 2024, each affecting over $1.5K$ files. LangChain shows the most dynamic evolution, with repeated churn peaks ($300K-400K$ lines) and a major deletion spike ($\sim400K$ lines) in mid-2025, indicating ongoing architectural restructuring.

\begin{figure}[ht]
    \centering
    \begin{subfigure}[t]{0.49\linewidth}
        \centering
        \includegraphics[width=\linewidth]{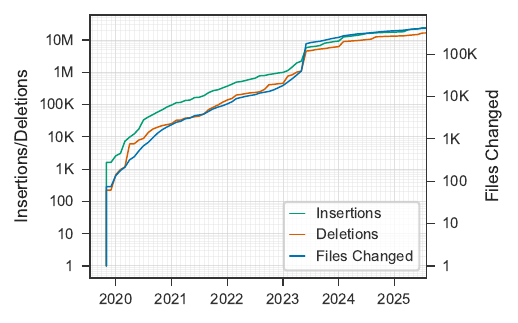}
        \label{fig:commits-cum-stats-trend}
    \end{subfigure}
    \hfill
    \begin{subfigure}[t]{0.49\linewidth}
        \centering
        \includegraphics[width=\linewidth]{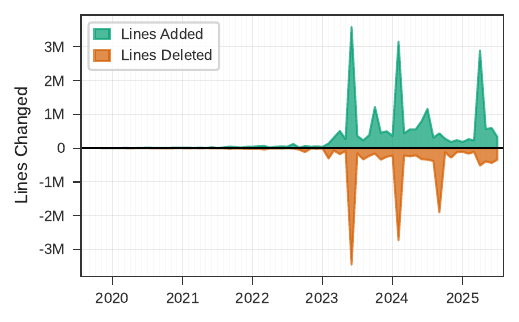}
        \label{fig:stacked_area_chart_commits_stats}
    \end{subfigure}
    \vspace{-1.5em}
    \caption{Ecosystem-level code evolution in multi-agent frameworks. (a) Cumulative growth shows overall expansion in code and file changes. (b) Temporal distribution highlights the balance between code additions and deletions over time.}
    \vspace{-2em}
    \label{fig:rq1-cumulative-stats}
\end{figure}

At the ecosystem level, Figure~\ref{fig:rq1-cumulative-stats}~(a) shows cumulative insertions and deletions reaching 10–20 million lines and over $100K$ files changed by 2025. The steepest growth occurred between 2020 and 2023. Figure~\ref{fig:rq1-cumulative-stats}~(b) highlights a shift toward maintainability: from 2023 onward, large-scale deletions were often balanced by equivalent insertions, suggesting increased focus on code restructuring rather than expansion.

\begin{myquote}
\textbf{Finding 1.1}. \textit{The \gls*{mas} framework ecosystem exhibits three distinct development profiles: sustained high-intensity development characterized by continuous elevated commit activity, steady consistent development with predictable patterns over time, and burst/sporadic development marked by concentrated periods of intense activity followed by minimal engagement. Most repositories intensified their development starting in 2023, suggesting a critical growth period for the ecosystem, though development consistency varies dramatically across projects with \glspl*{cv} ranging from 48.6\% to 456.1\%.}
\end{myquote}

% <<<<<<<<<< RQ 1.1

% RQ 1.2 >>>>>>>>>>
\subsubsection{What is the distribution of
commit types in MAS repositories?}

Characterizing the distribution of commit types provides valuable insights into the development of MAS. We used a fine-tuned DistilBERT model~\cite{hf-ml-msr}, trained on GitHub commit messages~\cite{sarwar2020multi}, to classify 42,266 commits in our dataset into three fundamental maintenance types: perfective commits that enhance system functionality, adaptive commits that address evolving requirements, and corrective commits that resolve defects and bugs, as well as their combinations.

As shown in Table~\ref{table:classification_commit}, perfective maintenance activities account for 40.83\% of all commits, which significantly exceeds both corrective~(27.36\%) and adaptive~(24.30\%) maintenance types. The three single-maintenance commits account for 92.49\% of all commits, while combined maintenance commits only account for 7.51\%.

Our analysis reveals distinct maintenance patterns across AI frameworks as shown in Table~\ref{tab:commit_classification_pct_frameworks}. Perfective maintenance dominates all frameworks~(34.2-51.5\%), indicating these systems prioritize feature development over corrective maintenance, characteristic of rapidly evolving domains. Semantic Kernel exhibits the highest perfective ratio~(51.5\%) with the lowest corrective maintenance~(18.3\%), suggesting a mature, stable codebase. Conversely, SuperAGI and Letta demonstrate higher corrective ratios~(32.8\% and 33.5\% respectively), potentially indicating less stable architectures or more experimental features.
Frameworks demonstrate considerable corrective and adaptive maintenance activity~(18.3-33.5\% and 17.2-27\% respectively), indicating active responses to both defects and changing requirements. 

\begin{table}[t]
    \centering
    \footnotesize
    \caption{Distribution of commit types across the entire dataset based on maintenance classification.}
    \vspace{-0.8em}
    \footnotesize
    \renewcommand{\arraystretch}{1.2}
    \setlength{\tabcolsep}{3pt}
    \begin{tabularx}{\columnwidth}{
        >{\raggedright\arraybackslash}X
        >{\raggedright\arraybackslash}X
    }
        \toprule
        \textbf{Commit Type} & \textbf{Percentage} \\
        \midrule
        Perfective & 40.83\% \\
        Corrective & 27.36\% \\
        Adaptive & 24.30\% \\
        Adaptive Perfective & 5.76\% \\
        Corrective Perfective & 1.36\% \\
        Corrective Adaptive & 0.39\% \\
        \bottomrule
    \end{tabularx}
    \vspace{-2em}
    \label{table:classification_commit}
\end{table}

\begin{table}
    \centering
    \footnotesize
    \caption{Distribution of commit classifications showing maintenance activity patterns across frameworks.\\ P: Perfective, C: Corrective, A: Adaptive, AP: Adaptive Perfective, CP: Corrective Perfective, CA: Corrective Adaptive}
    
    \label{tab:commit_classification_pct_frameworks}
    \renewcommand{\arraystretch}{1.2}
    \setlength{\tabcolsep}{3pt}
    \begin{tabularx}{\columnwidth}{>{\raggedright\arraybackslash}Xrrrrrr}
    \toprule
        \textbf{MAS} & \textbf{P} (\%) & \textbf{C} (\%) & \textbf{A} (\%) & \textbf{AP} (\%) & \textbf{CP} (\%) & \textbf{CA} (\%) \\
        \midrule
        AutoGen & 43.4 & 23.7 & 25.2 & 5.8 & 1.4 & 0.5 \\
        CrewAI & 44.2 & 24.5 & 24.0 & 6.0 & 0.9 & 0.5 \\
        Haystack & 42.6 & 26.4 & 24.3 & 5.5 & 0.9 & 0.4 \\
        LangChain & 39.2 & 26.9 & 25.3 & 6.8 & 1.3 & 0.4 \\
        Letta & 38.1 & 33.5 & 23.2 & 4.2 & 0.8 & 0.2 \\
        Llamaindex & 34.2 & 32.4 & 27.0 & 4.3 & 1.5 & 0.5 \\
        Semantic Kernel & 51.5 & 18.3 & 20.0 & 7.5 & 2.4 & 0.4 \\
        SuperAGI & 46.0 & 32.8 & 17.2 & 2.7 & 1.2 & 0.1 \\
        \bottomrule

\end{tabularx}
\vspace{-2em}
\end{table}

\begin{myquote}
\textbf{Finding 1.2}. \textit{The prevalence of perfective commits indicates that the development of MAS prioritizes continuous improvement over reactive bug fixing, indicating a proactive approach to software evolution. Mixed-category commits remain minimal~(\textless8\%), revealing atomic, single-purpose development practices where commits typically address focused maintenance tasks rather than combining multiple objectives. These patterns suggest the ecosystem is in an active growth phase, with development efforts concentrated on feature enhancement rather than traditional maintenance.} 
\end{myquote}
% <<<<<<<<<< RQ 1.2

\subsection{Issue Landscape (RQ2)}

% RQ 2.1 >>>>>>>>>>
\subsubsection{How do issue patterns evolve over time across frameworks?}
The \gls*{mas} repositories demonstrate heterogeneous issue reporting patterns with significant temporal variations in both volume and activity intensity.

\begin{figure}[ht]
    \centering
    \vspace{-0.3em}
    \begin{subfigure}[t]{0.49\linewidth}
        \centering
        \includegraphics[width=\linewidth]{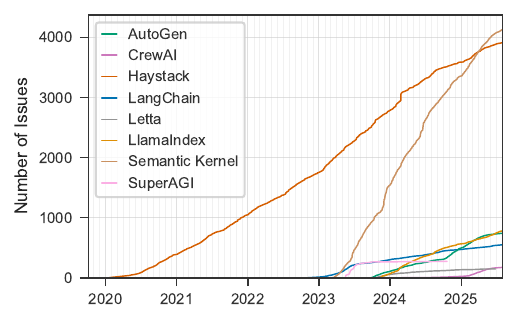}
        \label{fig:issue-cum-trend}
    \end{subfigure}
    \hfill
    \begin{subfigure}[t]{0.49\linewidth}
        \centering
        \includegraphics[width=\linewidth]{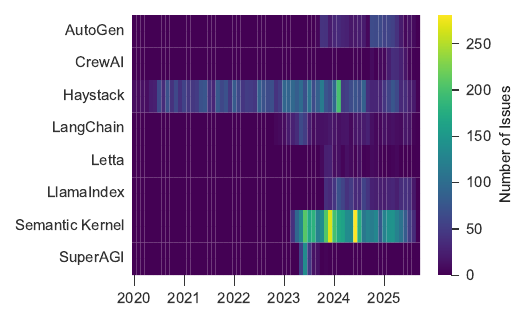}
        \label{fig:issues-heatmap}
    \end{subfigure}
    \vspace{-1.5em}
    \caption{Issue reporting activity across multi-agent frameworks. (a) Cumulative issue growth illustrates long-term reporting trends. (b) The monthly heatmap highlights bursts of issue creation and maintenance activity over time.}
    \vspace{-1em}
    \label{fig:rq2-issue-trends}
\end{figure}

As shown in Figure~\ref{fig:rq2-issue-trends}~(a), Haystack and Semantic Kernel emerge as the dominant repositories with approximately 4,000 cumulative issues each by 2025, while other frameworks maintain substantially lower volumes below 1,000 issues.
SuperAGI demonstrates a concentrated burst pattern with a dramatic spike in mid-2023 before declining sharply, contrasting with the more gradual emergence of LangChain, LlamaIndex, and other frameworks starting in 2023.

Figure~\ref{fig:rq2-issue-trends}~(b) further illustrates these patterns through temporal clustering, showing intense activity periods for Haystack and Semantic Kernel during 2023-2024 with values exceeding 200 issues per month, with Haystack exhibiting sustained high activity since 2020 and Semantic Kernel showing explosive growth starting in late 2023, while repositories like Letta and CrewAI display minimal activity.

\begin{figure}[t]
    \centering
    \includegraphics{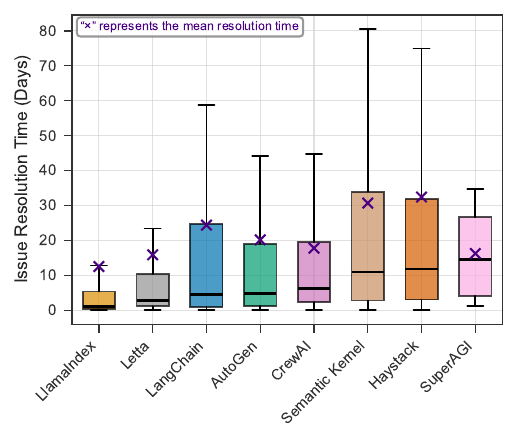}
    \vspace{-2.5em}
    \caption{Distribution of issue resolution times across repositories. Box plots show median (center line), \gls*{iqr} (box), whiskers extending to 1.5 × \gls*{iqr}, and mean values (×). Outliers are not shown.}
    \vspace{-2em}
    \label{fig:resolution-boxplot}
\end{figure}

Figure~\ref{fig:resolution-boxplot} reveals substantial variation in issue resolution efficiency across frameworks, with median resolution times ranging from approximately 1 day (LlamaIndex) to over 10 days (Semantic Kernel and Haystack). The boxplot whiskers extend to 1.5 times the \gls*{iqr}, indicating that outliers beyond this threshold are not displayed; thus, actual maximum resolution times may extend considerably beyond the visible range of approximately 80 days.
Notably, frameworks like LlamaIndex, Letta, and AutoGen show lower median resolution times with relatively compact \glspl*{iqr}, while Semantic Kernel and Haystack demonstrate higher medians and greater spread. Mean resolution times (marked with ``×'') consistently exceed medians across all frameworks, suggesting right-skewed distributions where a subset of issues requires substantially longer resolution times.

\begin{myquote}
\textbf{Finding 2.1}. \textit{The \gls*{mas} framework ecosystem exhibits highly heterogeneous issue reporting patterns with three distinct profiles: sustained high-volume activity (Haystack, Semantic Kernel with 4,000+ issues), moderate steady growth (LangChain, LlamaIndex with under 1,000 issues), and concentrated burst activity (SuperAGI). Issue reporting intensified dramatically across most frameworks starting in 2023, reflecting rapid ecosystem expansion. Resolution efficiency varies substantially across frameworks, with median times ranging from approximately 1 day to over 10 days. The consistent positioning of mean values above medians across all frameworks indicates right-skewed distributions where most issues resolve relatively quickly while a subset requires extended resolution times that may exceed the displayed range.}
\end{myquote}
% <<<<<<<<<< RQ 2.1

% RQ 2.2 >>>>>>>>>>
\subsubsection{What types of issues are most prevalent across frameworks?}
The analysis of normalized issue labels reveals a clear hierarchy of concern types across the \gls*{mas} framework ecosystem, with technical implementation challenges dominating over process management and community engagement.
\begin{table}
    \centering
    \caption{Distribution of issue labels across all repositories (an issue can have multiple labels).}
    \vspace{-0.8em}
    \renewcommand{\arraystretch}{1.2}
    \begin{tabularx}{\columnwidth}{XXX}
        \toprule
        \textbf{Label} & \textbf{Count} & \textbf{Percentage} \\
        \midrule
         Language/Framework  & 3,301     &     31\% \\
         Project Workflow  &   2,563     &     24\% \\
         Triage            &   2,439     &     23\% \\
         Bug               &   2,355     &     22\% \\
         Infrastructure    &   1,467     &     14\% \\
         Data Processing   &   1,161     &     11\% \\
         Agent Issues      &   1,049     &     10\% \\
         Documentation     &    734     &      7\% \\
         Feature           &    727     &      7\% \\
         Community         &    682     &      6\% \\
         Excluded          &    266     &      3\% \\
         User Experience   &    195     &      2\% \\
         Other             &    434     &      4\% \\
         \bottomrule
    \end{tabularx}
    \label{tab:issue-label-distribution}
    \vspace{-2em}
\end{table}

\begin{table*}
    \centering
    \caption{Description of the key features in the issues for MAS repositories.}
    \vspace{-1em}
    \renewcommand{\arraystretch}{1.2}
    % Disable hyphenation locally
    \begin{tabularx}{\textwidth}{>{\raggedright\arraybackslash}l>{\raggedright\arraybackslash}Xc}
        \toprule
        \textbf{Topic} & \textbf{Components} & \textbf{Frequency (\%)} \\
        \midrule
        Agent Framework Development & agent, declarative, framework, assistant, autogen, collaboration & 14.58\% \\ 
        Chat System \& Group Communication & chat, message, groupchat, user, agent, chathistory & 10.50\% \\ 
        Planning \& Sequential Execution & planner, stepwise, sequential, actionplanner, stepwiseplanner & 9.79\% \\ 
        AI Service Provider Integration & openai, plugin, api, sdk, endpoint, assistants & 9.48\% \\ 
        Evaluation \& Performance Metrics & evaluation, metric,  pipeline, evaluator, score, output, calculate & 8.97\% \\ 
        Model Training \& Fine-tuning & model, transformer, training, finetune, tinybert, gpu, distillation, cuda & 6.83\% \\ 
        Function Calling \& Exception Handling & functioncallingstepwiseplanner, throw, error, exception, functioncallcontent & 6.52\% \\ 
        Cloud Agent Orchestration & azureaiagent, azure, agent, streaming, message, thread, service & 5.81\% \\ 
        Prompt Engineering \& Token Management & prompt, handlebar, token, template, tokenizer, truncation, model, query, generate & 5.81\% \\ 
        Question Answering \& Text Processing & answer, question, table, text, document, generation, tableqa & 4.49\% \\ 
        Kernel Services \& Orchestration & kernel, semantic, mcp, openaiassistantagent, server, handofforchestration, orchestration & 4.38\% \\ 
        Documentation \& Search Features & page, documentation, search, matching, context, doc, bing, highlight, retrieval & 3.77\% \\ 
        Other & support, method, model, question, granularity, experiment, tool, image, test, implement & 9.07\% \\ 
        \bottomrule
    \end{tabularx}
    \vspace{-2em}
    \label{tbl:issue_topics}
\end{table*}
As shown in Table~\ref{tab:issue-label-distribution}, Language/Framework-specific tags represent the most frequent category at 31\%~(3,301 issues), though these labels (e.g., .NET, Java, Python) primarily serve as descriptive markers indicating the technology stack and are typically combined with other labels to provide meaningful context about the underlying issue type.
Project Workflow issues constitute 24\%~(2,563 issues) and Triage labels account for 23\%~(2,439 issues), collectively representing 47\% of all issue metadata; however, these categories reflect repository management processes (e.g., priority assignment, \gls*{pr} status, stale issue handling) rather than substantive product concerns, indicating substantial overhead in issue tracking and maintenance coordination.
Bug reports represent the most prevalent product-specific concern at 22\%~(2,355 issues), encompassing defects in functionality and unexpected behavior that require remediation.
Infrastructure issues account for 14\%~(1,467 issues), spanning build systems, deployment configurations, CI/CD pipelines, testing frameworks, dependencies, and platform-specific concerns (Docker, Windows, Azure), reflecting the operational complexity of maintaining \gls*{mas} frameworks across diverse environments.
Data Processing concerns constitute 11\%~(1,161 issues), addressing challenges in file conversion, indexing, metadata management, vector stores, and integration with various data backends (Elasticsearch, Pinecone, Weaviate), highlighting the centrality of data handling in agent workflows.

\begin{figure}[t]
    \centering
    \includegraphics{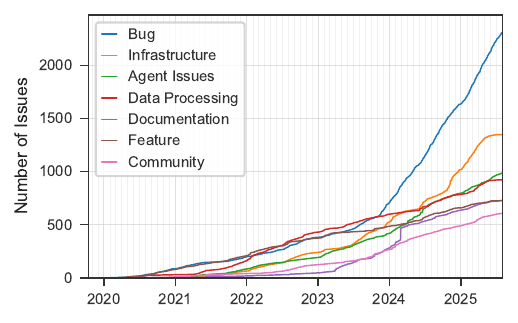}
    \vspace{-2em}
    \caption{Cumulative growth trajectories of major issue label categories}
    \vspace{-1em}
    \label{fig:label-cum-trend}
\end{figure}

Agent-related issues comprise 10\%~(1,049 issues), focusing specifically on agent behavior, multi-agent coordination, chat history management, planning mechanisms, function calling, and tool usage which is the core differentiating capabilities of \gls*{mas} frameworks.
Documentation and Feature requests each represent 7\%~(734 and 727 issues respectively), while Community engagement issues account for 6\%~(682 issues), encompassing help requests, good first issues, and contribution-seeking labels.
\Gls*{ux} concerns constitute only 2\%~(195 issues), covering \gls*{cli} interfaces, REST APIs, installation processes, and developer tooling, suggesting either robust \gls*{ux} design or underreporting of usability challenges.
Figure~\ref{fig:label-cum-trend} illustrates the cumulative growth trajectories of major issue categories, revealing that Bug reports have accumulated over 2,200 issues with accelerated growth starting in 2023, followed by Infrastructure~(approximately 1,400 issues) and Agent Issues~(approaching 1,000 issues), all showing similar inflection points around 2023 that coincide with the ecosystem's rapid expansion phase.
The high frequency of Bug~(22\%) and Infrastructure~(14\%) labels, combined with the relatively modest proportion of Agent-specific labels~(10\%), indicates that foundational software quality and operational stability challenges are reported as frequently or more frequently than issues related to the unique multi-agent capabilities these frameworks aim to provide.
Furthermore, the low prevalence of Documentation~(7\%), Community~(6\%), and \Gls*{ux}~(2\%) issues, totaling only 15\% collectively, suggests either effective support mechanisms in these areas or systematic underreporting of user-facing concerns relative to technical implementation problems.

\gls*{mas} introduce unique coordination and interaction challenges that differ from traditional software. Understanding the specific nature of these agent-related issues represents a core focus of our investigation.
We analyze agent issues in \gls*{mas} repositories, using BERTopic~\cite{grootendorst2022bertopic} to extract key discussion topics and validate their coherence through qualitative analysis.
Manual inspection of representative issues confirmed that BERTopic produced meaningful, semantically consistent clusters, revealing several issue categories listed in Table~\ref{tbl:issue_topics}.

The distribution indicates that Agent Systems \& Intelligence attracts the majority of developer attention~(58.42\%), encompassing Agent Framework Development~(14.58\%), Chat System \& Group Communication~(10.50\%), Planning \& Sequential Execution~(9.79\%), AI Service Provider Integration~(9.48\%), Cloud Agent Orchestration~(5.81\%), Question Answering \& Text Processing~(4.49\%), and Documentation \& Search Features~(3.77\%). Meanwhile, Technical Implementation \& Operations~(32.51\%) also receives substantial focus, underscoring the significant engineering challenges in deploying \glspl*{mas}. This category includes Evaluation \& Performance Metrics~(8.97\%), Model Training \& Fine-tuning~(6.83\%), Function Calling \& Exception Handling~(6.52\%), Prompt Engineering \& Token Management~(5.81\%), and Kernel Services \& Orchestration~(4.38\%).

\begin{myquote}
\textbf{Finding 2.2}. \textit{Technical concerns dominate the \gls*{mas} framework issue landscape, with Bugs~(22\%), Infrastructure~(14\%), Data Processing~(11\%), and Agent-specific issues~(10\%) as the most prevalent categories. Project management metadata appears frequently, indicating substantial maintenance overhead. All major categories show growth inflection points starting in 2023, correlating with ecosystem expansion. Topic analysis reveals developers prioritize agent capabilities~(58\%) while contending with operational challenges~(33\%), reflecting a feature-driven community facing persistent deployment barriers.}
\end{myquote}     
% <<<<<<<<<< RQ 2.2

%% file: threats.tex
\section{Threats to Validity}
\label{threats}
\stitle{Internal Validity.}
Potential threats to internal validity stem from the repository selection process. Although we selected the most active open-source multi-agent AI frameworks to ensure relevance, relying on GitHub metadata and project activity could still include peripheral or prototype repositories. To mitigate this, we applied filtering and cleaning steps to remove inactive forks and retain only repositories with consistent commit and issue activity. While this improves representativeness, some smaller yet meaningful projects may have been excluded.

\stitle{External Validity.}
Our findings are based solely on open-source projects hosted on GitHub. As a result, they may not fully generalize to proprietary multi-agent systems. Nevertheless, open-source frameworks dominate this emerging ecosystem, making GitHub an appropriate setting for studying early development and maintenance trends.

\stitle{Construct Validity.}
Metrics derived from commits and issues provide quantitative evidence of development and maintenance activity, but may overlook qualitative aspects such as code quality, design rationale, or developer expertise. We mitigated this limitation by combining multiple indicators, including commit types, code churn, issue categories, and responsiveness metrics, to strengthen construct alignment with real development practices.

\stitle{Reliability.}
Reliability depends on the reproducibility of our data collection and analysis procedures. Changes to GitHub’s API or repository structures could affect future replication. Automated classification of commit and issue text may also introduce minor labeling errors. To reduce these risks, we documented all data extraction and filtering steps and verified a subset of automated classifications for consistency.

By recognizing these limitations, we aim to ensure transparency about the study’s scope and provide confidence in the validity and reproducibility of our results.